\pdfoutput=1
\documentclass[%
 reprint,
superscriptaddress,
 amsmath,amssymb,
 aps,
 pra,
]{revtex4-2}

\usepackage{natbib}
\usepackage{graphicx}% Include figure files
\usepackage{dcolumn}% Align table columns on decimal point
\usepackage{bm}% bold math
\usepackage[usenames,dvipsnames]{color}
\usepackage{comment}
\begin{document}

\preprint{APS/123-QED}

\title{Quantum Interference of Distinguishable Photons Based on Spatially-Resolved Measurements}% Force line breaks with \\

\author{Miguel Angel Gonzalez}
\affiliation{Laboratorio de \'Optica Cu\'antica, Universidad de Los Andes, 4976 Bogot\'a, Colombia}
\author{Alejandra Alarcón}
\affiliation{Laboratorio de \'Optica Cu\'antica, Universidad de Los Andes, 4976 Bogot\'a, Colombia}
\author{Andres Camilo Quintero}
\affiliation{Laboratorio de \'Optica Cu\'antica, Universidad de Los Andes, 4976 Bogot\'a, Colombia}
\author{Daniel Sabogal}
\affiliation{Laboratorio de \'Optica Cu\'antica, Universidad de Los Andes, 4976 Bogot\'a, Colombia}
\author{Luca Maggio}
\affiliation{School of Mathematics and Physics, University of Portsmouth, Portsmouth PO1 3QL, United Kingdom}
\author{Vincenzo Tamma}
\affiliation{School of Mathematics and Physics, University of Portsmouth, Portsmouth PO1 3QL, United Kingdom}
\affiliation{Institute of Cosmology and Gravitation, University of Portsmouth, Portsmouth PO1 3FX, United Kingdom}
\author{Daniel F. Urrego}
\affiliation{ICFO – Institut de Ciencies Fotoniques, The Barcelona Institute of Science and Technology, 08860 Castelldefels, Barcelona, Spain}
\author{Alejandra Valencia}
\affiliation{Laboratorio de \'Optica Cu\'antica, Universidad de Los Andes, 4976 Bogot\'a, Colombia}

\date{\today}% It is always \today, today,

\begin{abstract}
We present experimental results demonstrating the quantum interference of two photons distinguishable in their transverse momenta, each entering the input ports of a balanced beam splitter. This counterintuitive interference effect is made possible through spatially resolved measurements in the near field, i.e., by resolving the conjugate variable in which the photons are distinguishable. Our experimental findings agree with theoretical predictions. We contrast our results with a non-spatially resolved measurement where averaging over the photons’ positions washes out the quantum interference observed in spatially resolved measurements.
\end{abstract}

\maketitle

Multiphoton interference, in its simplest form, refers to the scenario in which two photons, each entering a separate input port of a beam splitter, are jointly detected at its output ports. The most well-known example of this phenomenon is the observation of a dip in the coincidence count rate when the temporal delay between the incoming photons is zero~\cite{HOM, Shih-Alley-new}. This dip, commonly referred to as the Hong–Ou–Mandel (HOM) dip, is a signature of quantum interference between indistinguishable photons, a phenomenon that cannot be explained by classical physics. Beyond its foundational significance, multiphoton interference has led to the development of several practical applications. For example, quantum coherence tomography~\cite{CoherenceTomography-2015}, quantum sensing~\cite{Advances_Sensing,Tamma-sensing, salvatore-resolving, luca-resolving, Faccio_Sensing}, quantum metrology~\cite{Attosecond_HOM,Robust_Interferometer, Banaszek_Metrology}, and computation through Boson Sampling~\cite{aaronson-bosonsampling, tamma-multiboson, Tamma-Computational}.
\begin{figure}
\centering
\includegraphics[width=0.65\columnwidth]{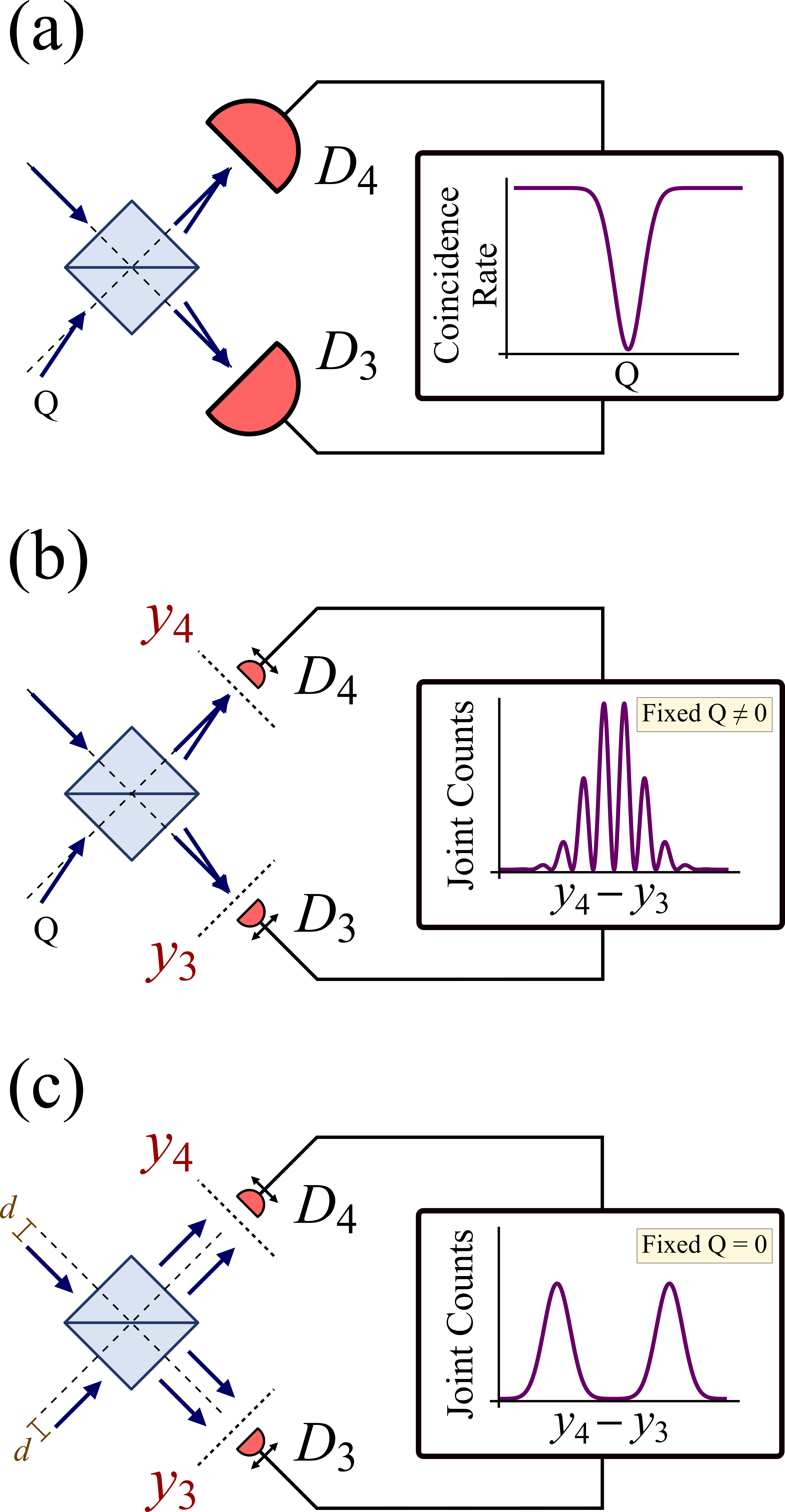}
\caption{Schematic representation of the key aspects for recovering spatial interference between photons distinguishable in transverse momentum. (a) Spatially non-resolved measurement: bucket-like detectors average over position, washing out spatial interference. A HOM dip is observed as a function of transverse momentum 
$Q$. (b) Spatially resolved measurement and observation in the appropriate variable: point-like detectors operate in the near-field (position basis), allowing interference to be recovered even when the photons are distinguishable in transverse momentum ($Q\neq0$). (c) Spatially resolved measurement illustrating the variable $d$ that tunes the distinguishability in the position variable.}
\label{cartoon}
\end{figure}

Achieving high-visibility interference in multiphoton experiments requires photons to be indistinguishable in all degrees of freedom. This poses a major challenge due to the scarcity of ideal single-photon sources that meet this condition~\cite{SinglePhoton_Mathias,Nature_SinglePhoton}. One strategy to overcome this is to perform resolved measurements in the conjugate degree of freedom to the one in which distinguishability arises. This approach has been successfully implemented in the time-frequency domain: interference between frequency-distinguishable photons has been recovered using time-resolved detection in  two-photon~\cite{rempe-quanbeats} and three-photon~\cite{Tamma-three} settings. Conversely, frequency-resolved measurements have been used to recover interference from temporally distinguishable photons~\cite{Tamma-freqresolved}, to exploit non-classical interference effects~\cite{Zou_Grayson_Mandel_1992, NIST_Japan, Mitchell}, and in optical coherence tomography~\cite{OCT}.

Light content is not restricted to the time-frequency degrees of freedom. In fact, light also has a rich structure when considering the transverse degrees of freedom. One well-known spatial property is orbital angular momentum (OAM), which has received significant attention in recent years~\cite{OAM-Padgett-2011, OAM-Padgett-2017}. Another important spatial variable is transverse position along with its conjugate, transverse momentum or spatial frequency. These variables have been experimentally studied in measurements revealing spatial HOM dips~\cite{Kim-spatial-HOM-fermion-boson, Lee-spatial-label, QuantumBeat_Momentum}, and in studies discussing non-classical interference~\cite{salvatore-resolving, luca-resolving, Ghosh_Mandel, Ou_Mandel_1989}. However, in contrast to the temporal domain, the spatial counterpart has not been as thoroughly investigated from the perspective of recovering quantum interference when photons are spatially distinguishable. In this context, with this paper, we aim to fill this gap by reporting experiments in which it is possible to relax the constraints imposed by photon distinguishability in spatial variables. Our strategy is to perform resolved measurements in the appropriate variable, i.e, the conjugate to the one in which there is distinguishability. This approach enables us to present a theoretical model and to experimentally observe interference in the transverse position domain, even when the photons are distinguishable in their transverse momenta. We also perform non-resolved measurements to highlight the role of resolved measurements to recover interference patterns that otherwise would be washed out.

To appreciate this approach, we clarify two aspects. First, the distinction between resolved and non-resolved measurements and second, the idea of measuring in the appropriate degree of freedom. Regarding the first aspect, lets begin by considering a spatial HOM experiment, as illustrated in Fig.~\ref{cartoon}(a). In this scheme, single photons enter the two input ports of a beam splitter and a difference in transverse momentum, $Q$, between them is introduced. This can be done, for example, by tilting the propagation direction of the photon entering port 2 relative to the one of the photon entering port 1. Two detectors, $D_3$ and $D_4$, positioned at the output ports, measure the coincidence rate $R_{\text{cc}}$ as a function of $Q$. This $R_{\text{cc}}(Q)$ is given by 
\begin{equation}
    R_{\text{cc}}(Q)= \int_0^Y G^{(2)}(y_3,y_4) \text{d}y_3\text{d}y_4,
    \label{roccpos}
\end{equation}
where $G^{(2)}(y_3, y_4)$ is the second-order spatial Glauber correlation function ~\cite{glauber-1963}, and $Y$ the spatial collection size (e.g., the pixel size). When $Y$ exceeds the spatial width of $G^{(2)}(y_3, y_4)$, the measurement averages over spatial features, acting as a bucket-like detector (exemplified by the big detectors in the cartoon), and the measurement for different values of $Q$ will show a HOM dip structure as illustrated in the graph in Fig.~\ref{cartoon}(a). This type of measurement leads to a spatially non-resolved measurement. In contrast, a measurement is spatially resolved when 
$Y$ is much smaller than the spatial width of $G^{(2)}(y_3, y_4)$ ~\cite{Masha-biphoton}, in fact, small enough to be treated as a point-like detectors (exemplified by the small detectors in Fig.~\ref{cartoon}(b)). Such measurements yield a histogram of joint detection events that provides direct access to the structure of $G^{(2)}(y_3, y_4)$.

The second aspect to clarify is the importance of measuring in the appropriate degree of freedom; specifically, in the variable conjugate to the one in which the photons are distinguishable. This concept originates from what Mandel referred to as the Alford and Gold effect~\cite{Mandel-alford-gold:62}. The key insight is that quantum interference can still be observed even when the photons are distinguishable in one variable, provided the measurement is made in its conjugate domain. For instance, if two optical pulses do not overlap in time, interference in the temporal domain disappears. However, when measured in the conjugate domain (frequency), interference fringes reappear. This effect has been experimentally demonstrated using laser light, both with pulses distinguishable in time that interfere in frequency~\cite{Salazar-Serrano:14}, and with laser beams distinguishable in position that interfere in transverse momentum~\cite{Flórez_2016}. In the multiphoton case, Fig.~\ref{cartoon}(b) illustrates a configuration where the Alford and Gold effect is observed in conjunction with spatially resolved measurements. Here, the photons are distinguishable in transverse momentum $Q$, yet interference, evidenced by the fringes in the graph, is recovered by performing position-resolved measurements, i.e, in the near-field regime.

When analyzing quantum interference in spatial variables, it is important to consider, in addition to $Q$, a variable $d$ that represents the lateral displacement of the photon path relative to the optical axis. This displacement is illustrated schematically in Fig.~\ref{cartoon}(c), where we set $Q=0$ to highlight the role of 
$d$ more clearly. A value of $d=0$ indicates that the photons are indistinguishable in position at the detection plane, which is precisely the case in Fig.~\ref{cartoon}(a) and (b). In contrast, Fig.~\ref{cartoon}(c) depicts a resolved measurement scenario and the situation in which $d\gg0$, i.e, photons that are distinguishable in position. In this case, no interference arises since the measurement is done in the same variable in which the photons are distinguishable. As a result, the joint detection pattern simply displays two lobes, reflecting the spatial intensity profile of the light emerging from the beam splitter.

To set the mathematical ground of our problem of interest, let's start by considering Fig.~\ref{cartoon}(b) and calculating the joint probability of detecting a photon at position $y_3$ and another photon at position $y_4$, $P_{\text{joint}}(y_3,y_4)$. This quantity is proportional to the spatial Glauber second order correlation function $G^{(2)}(y_3,y_4)$ given by
\begin{eqnarray}
\nonumber
G^{(2)}(y_3,y_4)&=&\langle1_1,1_2|\hat{\mathbf{E}}_3^{(-)}(y_3)\hat{\mathbf{E}}_4^{(-)}(y_4)\\
&&\hat{\mathbf{E}}_4^{(+)}(y_4)\hat{\mathbf{E}}_3^{(+)}(y_3)|1_1,1_2\rangle,
    \label{glauber}
\end{eqnarray}
with $\hat{\mathbf{E_j}}^{(-)}=( \hat{\mathbf{E_j}}^{(+)})^\dagger$ denoting the negative and positive parts of the electric field operators at detector $D_j$ with $j=\{3,4\}$. When dealing only with the spatial degree of freedom and considering the one dimensional case, each input photon can be described in the Fock state picture as a superposition of single mode states with transverse momentum $q$ weighted by a mode function $\zeta_i(q)$, i.e.,
\begin{equation}
  |1_i\rangle=\int_{-\infty}^{\infty}  \zeta_i(q)\hat{a}^{\dagger}_i (q)| 0\rangle\text{d}q,
  \label{state}
 \end{equation}
where the subindex $i=\{1,2\}$ denotes the input port of the beam splitter through which the photon is entering and $\hat{a}^{\dagger}_i(q)$ is the creation operator of a photon with transverse momentum $q$ in the input port $i$. Considering the case of both photons with equal polarization and therefore removing the need to account for the polarization vector, the positive part of the electric field at the detectors is
\begin{equation}
    \hat{E}_j^{(+)}(y_j)=\frac{1}{\sqrt{2\pi}}\int e^{-iqy_j}\hat{a}_j(q)\text{d}q.
    \label{electric field}
\end{equation}
By plugging Eq.~\eqref{state} and Eq.~\eqref{electric field} into Eq.~\eqref{glauber}, 
\begin{eqnarray}
\nonumber
    G^{(2)}(y_3,y_4)&=& |U_{1,3}U_{2,4}\chi_1(y_3)\chi_2(y_4)+\\
    &&U_{1,4}U_{2,3}\chi_1(y_4)\chi_2(y_3)|^2.
    \label{Pjoint_matrix}
\end{eqnarray}
Here $U_{i,j}$ denotes the elements of a matrix $U$ that represents the transformation done between the input and output modes because of the presence of the beam splitter. Mathematically, 
\begin{equation}
    \left(\begin{array}{c}\hat{a}_3\\ \hat{a}_4\end{array}\right)=U\left(\begin{array}{c}\hat{a}_1\\ \hat{a}_2\end{array}\right).
\end{equation}
The functions $\chi_i(y_j)$ are the spatial mode functions given by the Fourier transform of $\zeta_i(q)$, i.e., $\chi_i(y_j)=\frac{1}{\sqrt{2\pi}}\int_{-\infty}^{\infty}  e^{-iq_i y_j} \zeta_i(q_i) \text{d}q_i$.

By defining a matrix $\Lambda(y_3,y_4)$ such that
\begin{equation}
\Lambda(y_3,y_4)=\begin{pmatrix}
U_{1,3}\chi_1(y_3) & U_{1,4}\chi_1(y_4) \\
U_{2,3}\chi_2(y_3) & U_{2,4}\chi_2(y_4)
\end{pmatrix},
\end{equation}
the joint probability that satisfies $P_{\text{joint}}(y_3,y_4)\propto G^{(2)}(y_3,y_4)$ becomes
\begin{equation}
P_{\text{joint}}(y_3,y_4)\propto|\text{Perm}(\Lambda(y_3,y_4))|^2,
\end{equation}
where $\text{Perm}$ denotes the permanent of the matrix $\Lambda(y_3,y_4)$. This result is one of the key facts for which multiphoton interference has drawn attention in the areas of quantum computing and complexity, as it has been shown that this result can be extended to higher-dimensional matrices~\cite{tamma-multiboson}.

An analytical expression for $P_{\text{joint}}(y_3,y_4)$ can be obtained by defining the variable $Q=q_{20}-q_{10}$, where $q_{i0}$ denotes the central value of the transverse momentum distribution $\zeta_i(q)$, and considering a $50/50$ beam splitter as well as photons with a Gaussian spatial distribution. Specifically, the Gaussian distribution for each photon in the position variable can be written as
\begin{eqnarray}
\nonumber
\chi_{1}(y)&=&\left(\frac{2}{\pi  \text{w}_0^2}\right)^{1/4}\times\exp\left[-(y- d)^2/\text{w}_0^2\right]\\ &&\times\exp\left[i\left(q-\frac{Q}{2}\right)(y-d)\right] \label{chi1}\\
\nonumber
\end{eqnarray}
and
\begin{eqnarray}
\nonumber
\chi_{2}(y)&=&\left(\frac{2}{\pi  \text{w}_0^2}\right)^{1/4}\times\exp\left[-(y+ d)^2/\text{w}_0^2\right]\\ &&\times\exp\left[i\left(q+\frac{Q}{2}\right)(y+d)\right] \label{chi2},\\
\nonumber
\end{eqnarray}
with $\text{w}_0$ the waist of the spatial mode that quantifies the width of the spatial distribution as a standard deviation. By plugging Eq.~\eqref{chi1} and Eq.~\eqref{chi2} in Eq.~\eqref{Pjoint_matrix}, and setting $U$ as
\begin{equation}
    U=\frac{1}{\sqrt{2}}\begin{bmatrix}1 & 1 \\ 1 & -1 \end{bmatrix},
\end{equation}
which describes a $50/50$ beam splitter, the joint probability, for fixed values of $d$, $Q$ and $\text{w}_0$ becomes
\begin{eqnarray}
\nonumber
&&P_\text{joint}(y_3,y_4)=\frac{1}{\pi \text{w}_0^2}\exp\left[-\frac{2(y_3^2+y_4^2+2d^2)}{\text{w}_0^2}\right]\\
&&\times \left[\cosh\left(\frac{4d(y_4-y_3)}{\text{w}_0^2}\right)-\cos\left(Q(y_4-y_3)\right)\right].
\label{Pjoint_y3y4}
\end{eqnarray}

\begin{figure}
\centering\includegraphics[width=\columnwidth]{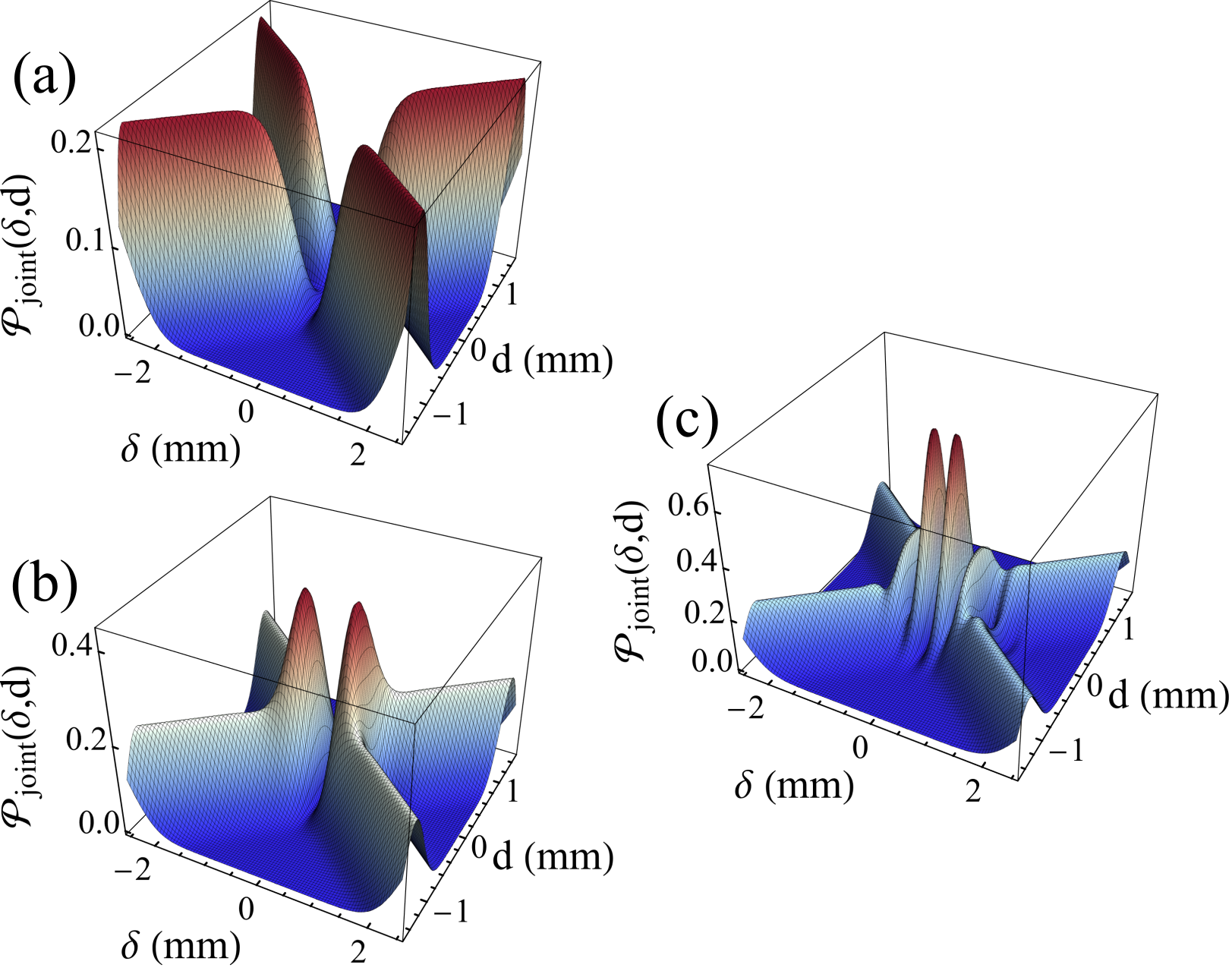}
\caption{Theoretical $\mathcal{P}_{\text{joint}}(\delta,d,Q,\text{w}_0)$ for different regimes of distinguishability in transverse momentum. Panel (a) shows the case for indistinguishable photons ($Q=0$) revealing total destructive interference for $d=0$, (b) for partially distinguishable photons ($Q=5~\text{mm}^{-1}$) and (c) for highly distinguishable photons ($Q=15~\text{mm}^{-1}$). The emergence of fringes reveals the appearance of interference in a spatially-resolved scheme. For all the graphs, $\text{w}_0=666~\mu\text{m}$}
\label{theory-insets}
\end{figure}

Since the joint probability is independent from a specific reference position point, Eq.~\eqref{Pjoint_y3y4} can be written in terms of the spatial difference between the detector's positions $\delta=y_4-y_3$. This is done by performing a change of variables, $y_3=y_0$ and $y_4=y_0+\delta$, and integrating over all possible values of $y_0$~\cite{leggero-teoria}. Going through this calculation leads to a joint probability that can be written as
\begin{eqnarray}
\nonumber
\mathcal{P}_{\text{joint}}(\delta)&=&\frac{1}{2\sqrt{\pi}\text{w}_0}\left[\cosh\left(\frac{4d\delta}{\text{w}_0^2}\right)-\cos\left(Q\delta\right)\right]\\
&&\exp\left(-\frac{4d^{2}+\delta^{2}}{\text{w}_0^{2}}\right).
\label{Pjoint_delta}
\end{eqnarray}

\begin{figure*}
\centering\includegraphics[scale= 0.25]{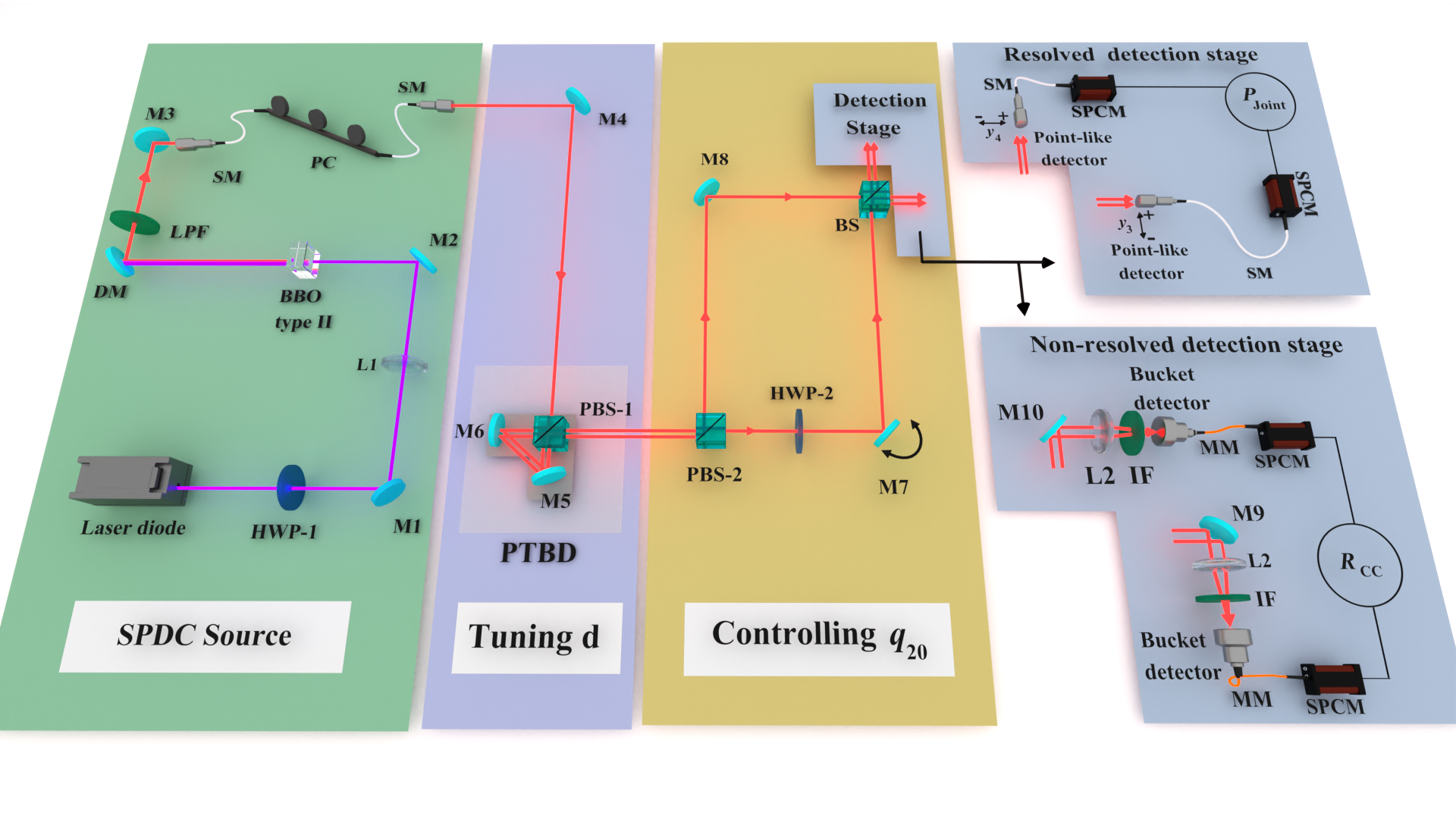}
\caption{Experimental setup to observe quantum interference of distinguishable photons in the transverse momentum variable. The two rightmost boxes depict a non-resolved detection scheme and a resolved detection scheme. Drawn with~\cite{drawingpage}.}
\label{setup_resolved}
\end{figure*}

Figure~\ref{theory-insets} illustrates the behavior of the joint probability described by Eq.~\eqref{Pjoint_delta} for different degrees of distinguishability in transverse momentum, quantified by the variable $Q$. In Fig.\ref{theory-insets}(a), where $Q=0$, the photons are completely indistinguishable in transverse momentum, and the absence of joint counts at $d=0$ signifies total destructive interference. Figure~\ref{theory-insets}(b), where $Q=5~\text{mm}^{-1}$, and Figure~\ref{theory-insets}(c), where $Q=15~\text{mm}^{-1}$, correspond to partially distinguishable and highly distinguishable photons, respectively. In these cases, the emergence of interference fringes is clearly visible, highlighting interference effects when a position-resolved measurement is done even for photons that are distinguishable in their transverse momentum.

The case of non-resolved detection can be obtained from Eq.~\eqref{Pjoint_delta} by integrating over all possible values of $\delta$. This allows to obtain the spatial rate of coincidence counts, $R_{\text{cc}}$, that becomes 
\begin{equation}
    R_{\text{cc}}(d,Q)=\frac{1}{2}\left[1-\exp\left( -\frac{16d^2+Q^2\text{w}_0^4}{4\text{w}_0^2}\right) \right].
    \label{Rocc}
\end{equation}
In Eq.~\eqref{Rocc}, all position-dependent information related to $y_3$ and $y_4$ is effectively erased as expected in a non-resolved measurement. $R_{\text{cc}}$ has a dependence on $d$ and $Q$. Taking $d=0$ in Eq.~\eqref{Rocc} gives as a result a spatial dip as a function of $Q$, which corresponds to the case depicted in Fig.~\ref{theory-insets}(a).
\begin{figure*}
\includegraphics[width=\textwidth]{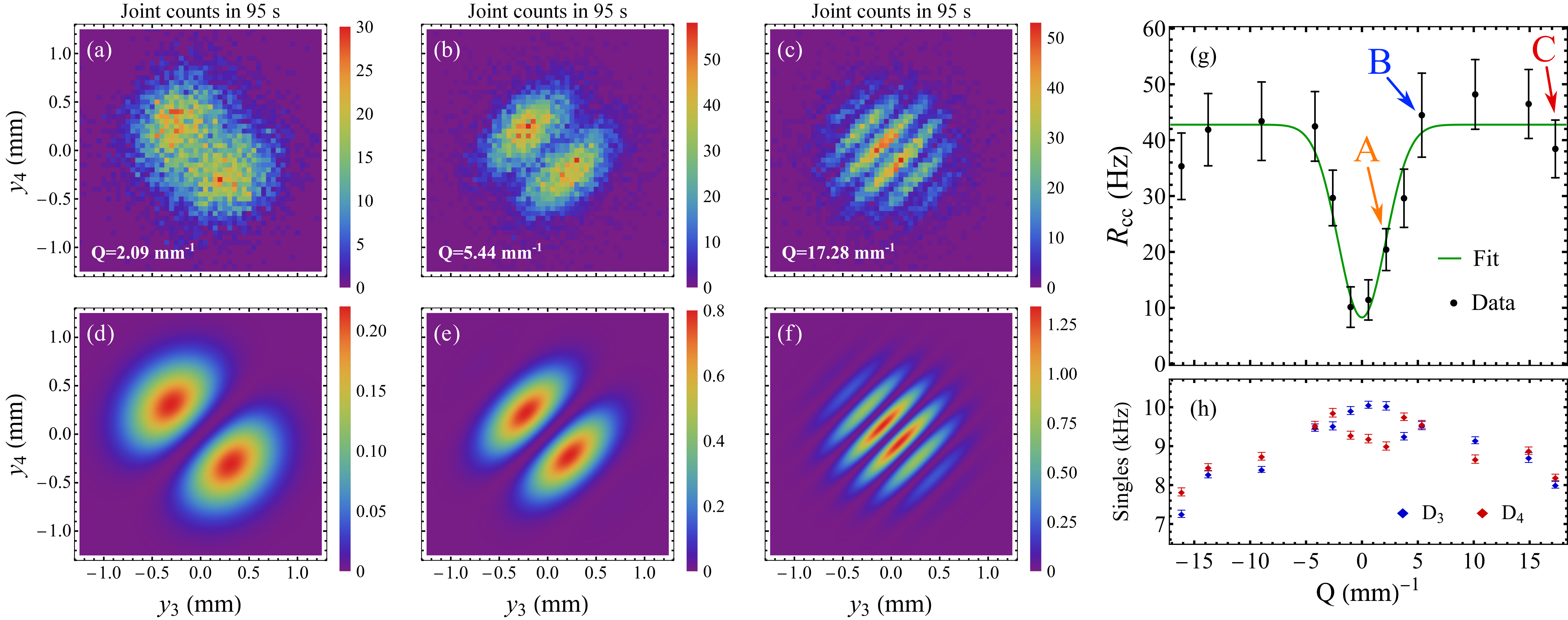}
\caption{Experimental demonstration of interference between photons that are distinguishable in transverse momentum. Agreement between theory and experiment is clear. (a-c) experimental data for $P_{\text{joint}}(y_3,y_4)$ for different degrees of distinguishability quantified by the value of $Q$ indicated in each figure. (d-f) theoretical model according to Eq.~\eqref{Pjoint_y3y4}. For both theory and experiment, $d=0~\mu\text{m}$ and $\text{w}_0=666~\mu\text{m}$. (g) experimental data for $R_\text{cc}(Q)$. Each point was obtained by averaging 60 one-second measurements in a 1.6~ns temporal coincidence window. Arrows $A$, $B$ and $C$ indicate the value of $Q$ used for the experimental data in (a), (b) and (c), respectively. Panel (h) depicts single counts at $D_3$ and $D_4$ as a function of $Q$. The error bars correspond to the standard deviation of each data point.}
\label{2d-Q-HOM}
\end{figure*}

To validate the theoretical predictions, we conducted three experiments using the setup illustrated in Fig.~\ref{setup_resolved}. The multiphoton interference described in Fig.~\ref{cartoon} takes place at the upper right corner of the setup, specifically at the beam splitter (BS) and the detection stage. The latter can be switched between resolved and non-resolved schemes, as illustrated by the rightmost boxes in Fig.~\ref{setup_resolved}. The components on the left side of the setup are essential for generating two independent photons that reach the BS, while also allowing precise control over the parameter $d$, and central transverse momentum, $q_{i0}$. The photons are created by means of Spontaneous Parametric Down-Conversion (SPDC) in a collinear 4~mm BBO type II crystal pumped by a 407~nm CW laser. The pump’s polarization state is adjusted using a half-wave plate (HWP-1) and the pump is focused into the crystal using a lens (L1), with focal length $f=100~mm$. After the crystal, the pump is removed using a dichroic mirror (DM) and a long-pass filter (LPF) with a cut-off wavelength of 750~nm. Due to the nature of SPDC, the photons from a pair are correlated in various degrees of freedom. For our experiment, the spatial correlations must be removed, and so the photons are coupled into a single-mode fiber (SM) that acts as a spatial filter. This SM is connected to a polarization controller (PC) that is used to correct the polarization state of the photons. After the PC, the spatial mode has a beam waist of approximately $\text{w}_0\approx666$~$\mu m$, leading to a Rayleigh range of $z_R\approx1.7~\text{m}$, ensuring that the light is collimated throughout its whole optical path during the experiment. Afterwards, the lateral displacement $d$ is induced by a polarization-based tunable beam displacer (PTBD) implemented with PBS-1 and mirrors, M5 and M6. The PTBD takes an input polarized beam and splits it into two parallel beams with orthogonal polarizations whose separation can be tuned \cite{Salazar-PTBD}. The photons are then sent through PBS-2, M7, M8 and the BS. This arrangement of optical elements directs each photon from an SPDC pair to a separate input port of the beam splitter, while ensuring identical arrival times to the BS. Half-wave plate 2 (HWP-2) is used to have the two SPDC photons with the same polarization: the polarization state of the horizontal photon is shifted to a vertical state. The control over the parameter $Q$ is done by changing the central transverse momentum $q_{20}$ by slightly tilting mirror M7 about the vertical axis using a gimbal mount. The resolution to estimate $Q$ is given by the rotation step size of this gimbal mount that leads to a minimum $Q$ of $0.8~\text{mm}^{-1}$.

For our first experiment, we measured $P_{\text{joint}}(y_3,y_4)$ using the detection scheme labeled ``resolved detection stage'' in  Fig.~\ref{setup_resolved}. It uses a rasterization technique based on point-like detectors in the near field, implemented with single-mode fibers conected to avalanche photodiodes 
$D_3$ and $D_4$. The input couplers of the single-mode fibers are mounted on motorized stages, enabling precise movement in both vertical and horizontal directions (minimum step $5~\mu\text{m}$). The vertical motion was merely used for alignment as we are only concerned with one-dimensional measurements done in the $y$-axis. For each pair of coordinates $(y_3,y_4)$, both individual events (singles) and joint counts were measured. The spatial resolution of the detection was defined by the detection collection mode, $\text{w}_R$, which was set to $\text{w}_R =50~\mu\text{m}$. This guarantees being within the spatially-resolved regime. The detectors were moved over a range of $[-1.25~\text{mm},1.25~\text{mm}]$ in steps of $50~\mu\text{m}$.

%Visibility:
%(a) V=75\pm1
%(b) V=66.7\pm0.5
%(c) V=64.1\pm0.7
%(g) V=82\pm7

%Fidelity:
%(a) F=0.814\pm0.004
%(b) F=0.907\pm0.002
%(c) F=0.884\pm0.003

%SD
%D_3: Average=9046.2 SD=858.712
%D_4: Average=8996.65 SD=614.592

The results for $P_{\text{joint}}(y_3,y_4)$, corresponding to three values of 
$Q$, associated to different degrees of distinguishability, are depicted in Fig.\ref{2d-Q-HOM}(a–c). The data clearly show that increasing distinguishability leads to the emergence of fringes, which become most pronounced in Fig.\ref{2d-Q-HOM}(c) which corresponds to highly distinguishable photons. This behavior demonstrates quantum interference between photons that are distinguishable in transverse momentum.
The specific values of $Q$ are depicted in each figure, and they were obtained by performing a fit of the experimental data to Eq.~\eqref{Pjoint_y3y4}, with $d=0$, and defining the visibility  $V$ as a parameter that multiplies the cosine term in Eq.~\eqref{Pjoint_y3y4}. From the fits, the visibilities were found to be $V\approx 75~\%$ (Fig.~\ref{2d-Q-HOM}(a)), $V\approx 67~\%$ (Fig.~\ref{2d-Q-HOM}(b)) and $V\approx 64~\%$ (Fig.~\ref{2d-Q-HOM}(c)). In the experiment, $Q$ is varied by rotating mirror M7. When this adjustment is made, the PTBD must be adjusted accordingly to keep $d$ as close to zero as the motor's step size allows ( $\sim 0.1\text{mm}$).

Figures.~\ref{2d-Q-HOM}(d-f) depict the theoretical model, according to Eq.~\eqref{Pjoint_y3y4} with $d=0$ and $\text{w}_0=666~\mu\text{m}$. An evident correspondence between theory and experiment is clearly seen. Indeed, the similarity between theory and experiment can be quantified by the similarity $S$ defined as~\cite{Fidelity_QuantumCoherence,Fidelity_QuantumWalks}
\begin{equation}
\label{fidelity}
S=\frac{\left(\sum_{y_3,y_4}\sqrt{P^\text{exp}_{\text{joint}}(y_3,y_4) P^\text{th}_{\text{joint}}(y_3,y_4)} \right)^2}{\sum_{y_3,y_4}P^\text{exp}_{\text{joint}}(y_3,y_4)\sum_{y_3,y_4}P^\text{th}_{\text{joint}}(y_3,y_4)},
\end{equation}
with the superscript $\text{exp}$ and $\text{th}$ refering to the experimental and theoretical joint distributions. The value of $S$ satisfies $0\leq S\leq1$, where a value of $S=1$ means that $P^\text{exp}_{\text{joint}}(y_3,y_4)=P^\text{th}_{\text{joint}}(y_3,y_4)$.  According to the results of Fig.~\ref{2d-Q-HOM}(a-f), for $Q=2.09~\text{mm}^{-1}$, $Q=5.44~\text{mm}^{-1}$ and $Q=17.28~\text{mm}^{-1}$, the values of $S$ are $S=0.81$, $S=0.91$ and $S=0.88$, respectively.

%$S=0.814\pm0.004$
%$S=0.907\pm0.002$
%$S=0.884\pm0.003$

For our second experiment, we employed the non-resolved detection scheme to contrast multiphoton interference between resolved and non-resolved measurements. Specifically, we measured a spatial HOM dip as a function of the transverse momentum variable $Q$, using the detection configuration illustrated in the part of Fig.~\ref{setup_resolved}, labeled ``non-resolved detection stage.'' For this, we inserted flip-mirrors M9 and M10 and used two bucket detectors, each constructed with a lens (L2) of focal length $f=200~\text{mm}$ and multimod e fibers. Additionally, narrow-bandpass filters, IF, (Andover 050FC14-25/814.0-D) were placed in front of the detectors to enhance the visibility of the interference pattern. The experimental results are shown as points in Fig.\ref{2d-Q-HOM}(g). The presence of a dip in the rate of coincidence counts is clear. The solid line corresponds to a fit of the data to Eq.~\eqref{Rocc}. For this, we set $d$ and $\text{w}_0$ as indicated in the figure caption and used as fitting parameter a visibility, $V$, that multiplies the exponential term (see Supplementary Material). With this fit, a visibility of $V \approx 81 \%$ was found for our experimental data. Figure~\ref{2d-Q-HOM}(h) shows the single counts recorded at each detector. These counts remain nearly constant, fluctuating by about $\sim 800$ Hz around an average of approximately $9$~kHz. This stability indicates the absence of first-order interference, confirming that the observed dip arises from genuine quantum interference. The drop in single counts at the edges of the graph is attributed to limitations in fiber coupling efficiency.

\begin{figure}
\centering
\includegraphics[width=\columnwidth]{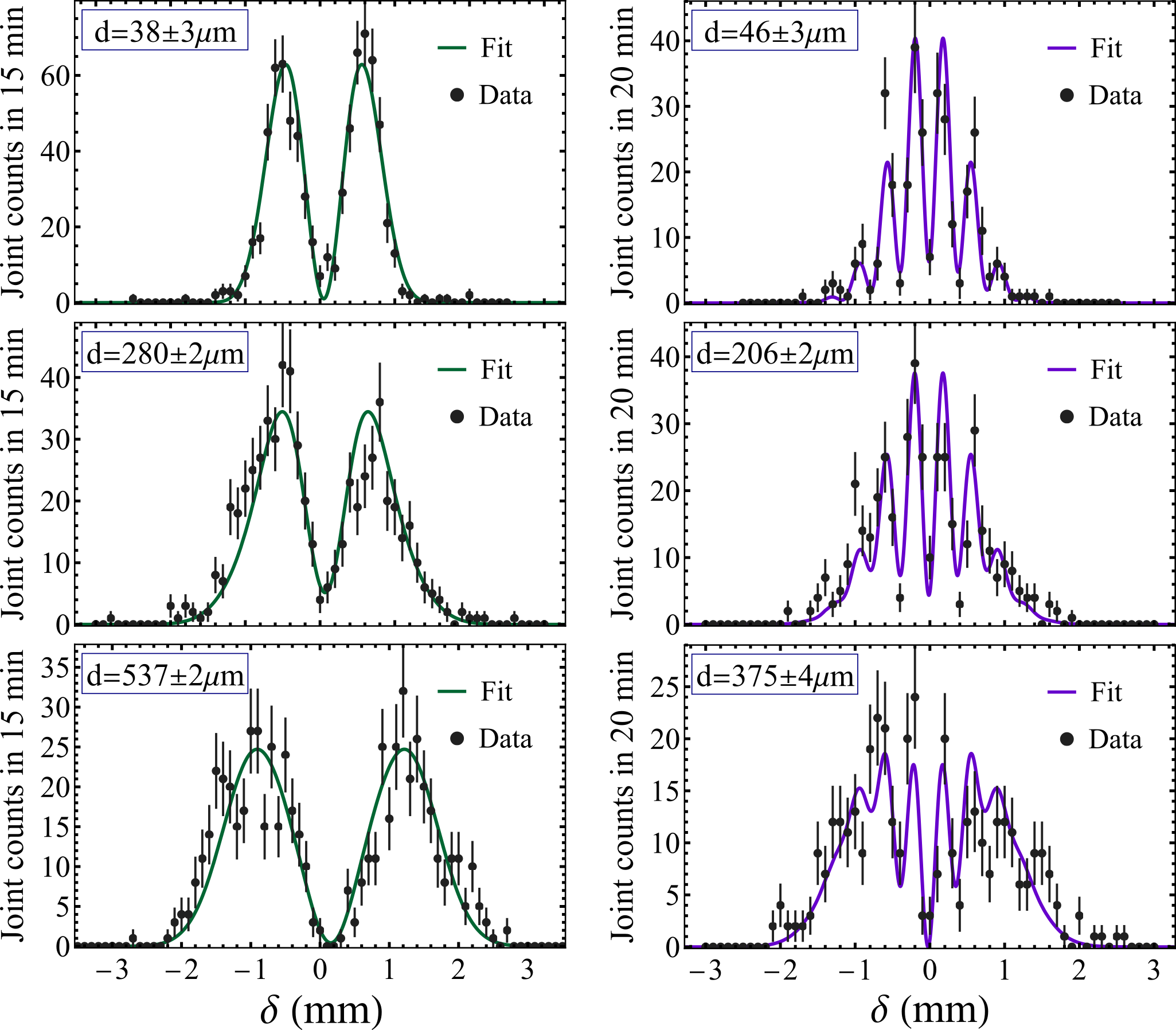}
\caption{Experimental data (dots) and fit to theoretical model (solid lines) for $\mathcal{P}_\text{joint}(\delta,d,Q,\text{w}_0)$ at different values of $d$ for two degrees of distinguishability in transverse momentum. Left column depicts the case for partially distinguishable ($Q\approx4.19~\text{mm}^{-1}$) and right column depicts the case for highly distinguishable ($Q\approx16.28~\text{mm}^{-1}$) photons. The emerging of fringes as expected by observing Fig.~\ref{theory-insets} is evident. For these measurements, the waist was kept at $\text{w}_0=666~\mu\text{m}$. }
\label{3d-Q}
\end{figure}

The arrows in Fig.~\ref{2d-Q-HOM}(g) indicate the points named $A$, $B$, and $C$ that correspond to the values of $Q$ in which the resolved measurements of Figs.~\ref{2d-Q-HOM}(a-c) were performed. There are two important issues to note. On the one hand, for points in the flat region of the HOM dip (rightmost blue (B) and red arrows (C)), i.e., where the photons are distinguishable in transverse momentum and do not reveal interference in non-resolved measurements, a clear interference pattern appears when resolved measurements are done. On the other hand, for points closer to $Q=0$ (yellow arrow (A)) in the non-resolved case, the interference manifests itself by a drastic decrease in $R_{\text{cc}}(Q)$, while the resolved measurements do not reveal the presence of fringes due to the effect of destructive interference. 

%Visibilities:
%(a) V=96\pm2
%(b) V=80\pm7
%(c) V=90\pm10

%(d) V=75\pm7
%(e) V=80\pm9
%(f) V=95\pm5

Since our setup provides full simultaneous tunability over $d$ and $Q$, we performed a third experiment to partially measure the graphs seen in Fig.~\ref{theory-insets}. This was done by performing measurements in the variable $\delta$ for different values of $d$. Operationally, this means setting a fixed value for $d$ and moving the detectors in tandem ($y_3$ going from positive to negative and $y_4$ from negative to positive) while ensuring that the coordinate pairs obey the relation $y_3+y_4=0$. Since the number of data points taken in these measurements was considerably lower than the ones shown in Fig.~\ref{2d-Q-HOM}(a-c), we placed the same narrow-bandpass filters, as in the non-resolved measurement, in front of the detectors that allowed us to increase the visibility of interference. The results are depicted as dots in Fig.~\ref{3d-Q}, for the case of partially distinguishable photons (left column) and
highly distinguishable photons (right column). In both cases, three different values of $d$ were chosen. For distinguishable photons, two lobes appear that become wider and lower as $d$ increases. On the other hand, for distinguishable photons in transverse momentum, the presence of the fringes is clear, and its variation with respect to $d$ goes according to the theory depicted in Fig.~\ref{theory-insets}(c). In both columns, destructive interference is revealed as a notorious drop of the joint counts for $\delta=0$. The solid line in Fig.~\ref{3d-Q} corresponds to fitting the data to Eq.~\eqref{Pjoint_delta}. For this fit, we introduced a visibility, $V$, that multiplies the cosine term (see Supplementary Material) and used the values of $Q$, $d$ and $\text{w}_0$ indicated in Fig.~\ref{3d-Q} and in its caption. The fit leads to  visibilities ranging from $V=75\%$ to $V=96\%$.

In conclusion, we have experimentally demonstrated spatial quantum interference between two photons incident on a beam splitter, with tunable distinguishability in transverse momentum. Interference fringes observed under spatially resolved detection confirm that quantum interference can persist in the position domain, even when photons are distinguishable in transverse momentum. Additionally, a HOM-type dip was observed using non-resolved spatial measurements, underscoring the critical role of measurement resolution in recovering interference. These results illustrate that resolved measurements in the appropriate degree of freedom can effectively overcome the limitations imposed by photon distinguishability. Our work broadens the landscape of accessible degrees of freedom for quantum interference and its applications by extending multiphoton interference to spatial variables, specifically position and transverse momentum.

This project was supported by Facultad de Ciencias, Universidad de los Andes (INV-2022-143-2525, INV-2022-151-2625, INV-2023-163-2753, INV-2023-162-2836, INV-2024-188-3108). LM acknowledges partial support by Xairos Systems Inc. VT acknowledges support from the Air Force office of Scientific Research under award number FA8655-23-1- 7046.

%\bmsection{Disclosures}
%The authors declare no conflicts of interest.

%\bmsection{Data Availability Statement}
%Data underlying the results presented in this paper are not publicly available but may be obtained from the corresponding author upon reasonable request.

%%%%%%%%%%%%%%%%%%%%%%% References %%%%%%%%%%%%%%%%%%%%%%%%%

%\nocite{*}

\bibliography{apssamp}% Produces the bibliography via BibTeX.

\begin{thebibliography}{40}%
\makeatletter
\providecommand \@ifxundefined [1]{%
 \@ifx{#1\undefined}
}%
\providecommand \@ifnum [1]{%
 \ifnum #1\expandafter \@firstoftwo
 \else \expandafter \@secondoftwo
 \fi
}%
\providecommand \@ifx [1]{%
 \ifx #1\expandafter \@firstoftwo
 \else \expandafter \@secondoftwo
 \fi
}%
\providecommand \natexlab [1]{#1}%
\providecommand \enquote  [1]{``#1''}%
\providecommand \bibnamefont  [1]{#1}%
\providecommand \bibfnamefont [1]{#1}%
\providecommand \citenamefont [1]{#1}%
\providecommand \href@noop [0]{\@secondoftwo}%
\providecommand \href [0]{\begingroup \@sanitize@url \@href}%
\providecommand \@href[1]{\@@startlink{#1}\@@href}%
\providecommand \@@href[1]{\endgroup#1\@@endlink}%
\providecommand \@sanitize@url [0]{\catcode `\\12\catcode `\$12\catcode `\&12\catcode `\#12\catcode `\^12\catcode `\_12\catcode `\%12\relax}%
\providecommand \@@startlink[1]{}%
\providecommand \@@endlink[0]{}%
\providecommand \url  [0]{\begingroup\@sanitize@url \@url }%
\providecommand \@url [1]{\endgroup\@href {#1}{\urlprefix }}%
\providecommand \urlprefix  [0]{URL }%
\providecommand \Eprint [0]{\href }%
\providecommand \doibase [0]{https://doi.org/}%
\providecommand \selectlanguage [0]{\@gobble}%
\providecommand \bibinfo  [0]{\@secondoftwo}%
\providecommand \bibfield  [0]{\@secondoftwo}%
\providecommand \translation [1]{[#1]}%
\providecommand \BibitemOpen [0]{}%
\providecommand \bibitemStop [0]{}%
\providecommand \bibitemNoStop [0]{.\EOS\space}%
\providecommand \EOS [0]{\spacefactor3000\relax}%
\providecommand \BibitemShut  [1]{\csname bibitem#1\endcsname}%
\let\auto@bib@innerbib\@empty
%</preamble>
\bibitem [{\citenamefont {Hong}\ \emph {et~al.}(1987)\citenamefont {Hong}, \citenamefont {Ou},\ and\ \citenamefont {Mandel}}]{HOM}%
  \BibitemOpen
  \bibfield  {author} {\bibinfo {author} {\bibfnamefont {C.~K.}\ \bibnamefont {Hong}}, \bibinfo {author} {\bibfnamefont {Z.~Y.}\ \bibnamefont {Ou}},\ and\ \bibinfo {author} {\bibfnamefont {L.}~\bibnamefont {Mandel}},\ }\bibfield  {title} {\bibinfo {title} {Measurement of subpicosecond time intervals between two photons by interference},\ }\href {https://doi.org/10.1103/PhysRevLett.59.2044} {\bibfield  {journal} {\bibinfo  {journal} {Phys. Rev. Lett.}\ }\textbf {\bibinfo {volume} {59}},\ \bibinfo {pages} {2044} (\bibinfo {year} {1987})}\BibitemShut {NoStop}%
\bibitem [{\citenamefont {Shih}\ and\ \citenamefont {Alley}(1988)}]{Shih-Alley-new}%
  \BibitemOpen
  \bibfield  {author} {\bibinfo {author} {\bibfnamefont {Y.~H.}\ \bibnamefont {Shih}}\ and\ \bibinfo {author} {\bibfnamefont {C.~O.}\ \bibnamefont {Alley}},\ }\bibfield  {title} {\bibinfo {title} {New type of einstein-podolsky-rosen-bohm experiment using pairs of light quanta produced by optical parametric down conversion},\ }\href {https://doi.org/10.1103/PhysRevLett.61.2921} {\bibfield  {journal} {\bibinfo  {journal} {Phys. Rev. Lett.}\ }\textbf {\bibinfo {volume} {61}},\ \bibinfo {pages} {2921} (\bibinfo {year} {1988})}\BibitemShut {NoStop}%
\bibitem [{\citenamefont {Okano}\ \emph {et~al.}(2015)\citenamefont {Okano}, \citenamefont {Lim}, \citenamefont {Okamoto}, \citenamefont {Nishizawa}, \citenamefont {Kurimura},\ and\ \citenamefont {Takeuchi}}]{CoherenceTomography-2015}%
  \BibitemOpen
  \bibfield  {author} {\bibinfo {author} {\bibfnamefont {M.}~\bibnamefont {Okano}}, \bibinfo {author} {\bibfnamefont {H.~H.}\ \bibnamefont {Lim}}, \bibinfo {author} {\bibfnamefont {R.}~\bibnamefont {Okamoto}}, \bibinfo {author} {\bibfnamefont {N.}~\bibnamefont {Nishizawa}}, \bibinfo {author} {\bibfnamefont {S.}~\bibnamefont {Kurimura}},\ and\ \bibinfo {author} {\bibfnamefont {S.}~\bibnamefont {Takeuchi}},\ }\bibfield  {title} {\bibinfo {title} {0.54 $\mu$m resolution two-photon interference with dispersion cancellation for quantum optical coherence tomography},\ }\href {https://doi.org/10.1038/srep18042} {\bibfield  {journal} {\bibinfo  {journal} {Sci. Rep.}\ }\textbf {\bibinfo {volume} {5}},\ \bibinfo {pages} {18042} (\bibinfo {year} {2015})}\BibitemShut {NoStop}%
\bibitem [{\citenamefont {Pirandola}\ \emph {et~al.}(2018)\citenamefont {Pirandola}, \citenamefont {Bardhan}, \citenamefont {Gehring}, \citenamefont {Weedbrook},\ and\ \citenamefont {Lloyd}}]{Advances_Sensing}%
  \BibitemOpen
  \bibfield  {author} {\bibinfo {author} {\bibfnamefont {S.}~\bibnamefont {Pirandola}}, \bibinfo {author} {\bibfnamefont {B.~R.}\ \bibnamefont {Bardhan}}, \bibinfo {author} {\bibfnamefont {T.}~\bibnamefont {Gehring}}, \bibinfo {author} {\bibfnamefont {C.}~\bibnamefont {Weedbrook}},\ and\ \bibinfo {author} {\bibfnamefont {S.}~\bibnamefont {Lloyd}},\ }\bibfield  {title} {\bibinfo {title} {Advances in photonic quantum sensing},\ }\href {https://doi.org/10.1038/s41566-018-0301-6} {\bibfield  {journal} {\bibinfo  {journal} {Nat. Photonics}\ }\textbf {\bibinfo {volume} {12}},\ \bibinfo {pages} {724} (\bibinfo {year} {2018})}\BibitemShut {NoStop}%
\bibitem [{\citenamefont {Triggiani}\ and\ \citenamefont {Tamma}(2024)}]{Tamma-sensing}%
  \BibitemOpen
  \bibfield  {author} {\bibinfo {author} {\bibfnamefont {D.}~\bibnamefont {Triggiani}}\ and\ \bibinfo {author} {\bibfnamefont {V.}~\bibnamefont {Tamma}},\ }\bibfield  {title} {\bibinfo {title} {Estimation with ultimate quantum precision of the transverse displacement between two photons via two-photon interference sampling measurements},\ }\href {https://doi.org/10.1103/PhysRevLett.132.180802} {\bibfield  {journal} {\bibinfo  {journal} {Phys. Rev. Lett.}\ }\textbf {\bibinfo {volume} {132}},\ \bibinfo {pages} {180802} (\bibinfo {year} {2024})}\BibitemShut {NoStop}%
\bibitem [{\citenamefont {Muratore}\ \emph {et~al.}(2025)\citenamefont {Muratore}, \citenamefont {Triggiani},\ and\ \citenamefont {Tamma}}]{salvatore-resolving}%
  \BibitemOpen
  \bibfield  {author} {\bibinfo {author} {\bibfnamefont {S.}~\bibnamefont {Muratore}}, \bibinfo {author} {\bibfnamefont {D.}~\bibnamefont {Triggiani}},\ and\ \bibinfo {author} {\bibfnamefont {V.}~\bibnamefont {Tamma}},\ }\bibfield  {title} {\bibinfo {title} {Superresolution imaging of two incoherent sources via two-photon-interference sampling measurements of the transverse momenta},\ }\href@noop {} {\bibfield  {journal} {\bibinfo  {journal} {Physical Review Applied}\ }\textbf {\bibinfo {volume} {23}},\ \bibinfo {pages} {054033} (\bibinfo {year} {2025})}\BibitemShut {NoStop}%
\bibitem [{\citenamefont {Maggio}\ and\ \citenamefont {Tamma}(2025)}]{luca-resolving}%
  \BibitemOpen
  \bibfield  {author} {\bibinfo {author} {\bibfnamefont {L.}~\bibnamefont {Maggio}}\ and\ \bibinfo {author} {\bibfnamefont {V.}~\bibnamefont {Tamma}},\ }\bibfield  {title} {\bibinfo {title} {Ultimate quantum sensitivity in the 3d relative localisation of two single-photon emitters via two-photon interference},\ }\href@noop {} {\bibfield  {journal} {\bibinfo  {journal} {arXiv preprint arXiv:2504.16294}\ } (\bibinfo {year} {2025})}\BibitemShut {NoStop}%
\bibitem [{\citenamefont {Ndagano}\ \emph {et~al.}(2022)\citenamefont {Ndagano}, \citenamefont {Defienne}, \citenamefont {Branford}, \citenamefont {Shah}, \citenamefont {Lyons}, \citenamefont {Westerberg}, \citenamefont {Gauger},\ and\ \citenamefont {Faccio}}]{Faccio_Sensing}%
  \BibitemOpen
  \bibfield  {author} {\bibinfo {author} {\bibfnamefont {B.}~\bibnamefont {Ndagano}}, \bibinfo {author} {\bibfnamefont {H.}~\bibnamefont {Defienne}}, \bibinfo {author} {\bibfnamefont {D.}~\bibnamefont {Branford}}, \bibinfo {author} {\bibfnamefont {Y.~D.}\ \bibnamefont {Shah}}, \bibinfo {author} {\bibfnamefont {A.}~\bibnamefont {Lyons}}, \bibinfo {author} {\bibfnamefont {N.}~\bibnamefont {Westerberg}}, \bibinfo {author} {\bibfnamefont {E.~M.}\ \bibnamefont {Gauger}},\ and\ \bibinfo {author} {\bibfnamefont {D.}~\bibnamefont {Faccio}},\ }\bibfield  {title} {\bibinfo {title} {Quantum microscopy based on hong--ou--mandel interference},\ }\href {https://doi.org/10.1038/s41566-022-00980-6} {\bibfield  {journal} {\bibinfo  {journal} {Nat. Photonics}\ }\textbf {\bibinfo {volume} {16}},\ \bibinfo {pages} {384} (\bibinfo {year} {2022})}\BibitemShut {NoStop}%
\bibitem [{\citenamefont {Lyons}\ \emph {et~al.}(2018)\citenamefont {Lyons}, \citenamefont {Knee}, \citenamefont {Bolduc}, \citenamefont {Roger}, \citenamefont {Leach}, \citenamefont {Gauger},\ and\ \citenamefont {Faccio}}]{Attosecond_HOM}%
  \BibitemOpen
  \bibfield  {author} {\bibinfo {author} {\bibfnamefont {A.}~\bibnamefont {Lyons}}, \bibinfo {author} {\bibfnamefont {G.~C.}\ \bibnamefont {Knee}}, \bibinfo {author} {\bibfnamefont {E.}~\bibnamefont {Bolduc}}, \bibinfo {author} {\bibfnamefont {T.}~\bibnamefont {Roger}}, \bibinfo {author} {\bibfnamefont {J.}~\bibnamefont {Leach}}, \bibinfo {author} {\bibfnamefont {E.~M.}\ \bibnamefont {Gauger}},\ and\ \bibinfo {author} {\bibfnamefont {D.}~\bibnamefont {Faccio}},\ }\bibfield  {title} {\bibinfo {title} {Attosecond-resolution hong-ou-mandel interferometry},\ }\href {https://doi.org/10.1126/sciadv.aap9416} {\bibfield  {journal} {\bibinfo  {journal} {Sci. Adv.}\ }\textbf {\bibinfo {volume} {4}},\ \bibinfo {pages} {eaap9416} (\bibinfo {year} {2018})}\BibitemShut {NoStop}%
\bibitem [{\citenamefont {Aguilar}\ \emph {et~al.}(2020)\citenamefont {Aguilar}, \citenamefont {Piera}, \citenamefont {Saldanha}, \citenamefont {Filho},\ and\ \citenamefont {Walborn}}]{Robust_Interferometer}%
  \BibitemOpen
  \bibfield  {author} {\bibinfo {author} {\bibfnamefont {G.~H.}\ \bibnamefont {Aguilar}}, \bibinfo {author} {\bibfnamefont {R.~S.}\ \bibnamefont {Piera}}, \bibinfo {author} {\bibfnamefont {P.~L.}\ \bibnamefont {Saldanha}}, \bibinfo {author} {\bibfnamefont {R.~L. d.~M.}\ \bibnamefont {Filho}},\ and\ \bibinfo {author} {\bibfnamefont {S.~P.}\ \bibnamefont {Walborn}},\ }\bibfield  {title} {\bibinfo {title} {Robust interferometric sensing using two-photon interference},\ }\href {https://doi.org/10.1103/PhysRevApplied.14.024028} {\bibfield  {journal} {\bibinfo  {journal} {Phys. Rev. Appl.}\ }\textbf {\bibinfo {volume} {14}},\ \bibinfo {pages} {024028} (\bibinfo {year} {2020})}\BibitemShut {NoStop}%
\bibitem [{\citenamefont {Chrapkiewicz}\ \emph {et~al.}(2016)\citenamefont {Chrapkiewicz}, \citenamefont {Jachura}, \citenamefont {Banaszek},\ and\ \citenamefont {Wasilewski}}]{Banaszek_Metrology}%
  \BibitemOpen
  \bibfield  {author} {\bibinfo {author} {\bibfnamefont {R.}~\bibnamefont {Chrapkiewicz}}, \bibinfo {author} {\bibfnamefont {M.}~\bibnamefont {Jachura}}, \bibinfo {author} {\bibfnamefont {K.}~\bibnamefont {Banaszek}},\ and\ \bibinfo {author} {\bibfnamefont {W.}~\bibnamefont {Wasilewski}},\ }\bibfield  {title} {\bibinfo {title} {Hologram of a single photon},\ }\href {https://doi.org/10.1038/nphoton.2016.129} {\bibfield  {journal} {\bibinfo  {journal} {Nat. Photonics}\ }\textbf {\bibinfo {volume} {10}},\ \bibinfo {pages} {576} (\bibinfo {year} {2016})}\BibitemShut {NoStop}%
\bibitem [{\citenamefont {Aaronson}\ and\ \citenamefont {Arkhipov}(2011)}]{aaronson-bosonsampling}%
  \BibitemOpen
  \bibfield  {author} {\bibinfo {author} {\bibfnamefont {S.}~\bibnamefont {Aaronson}}\ and\ \bibinfo {author} {\bibfnamefont {A.}~\bibnamefont {Arkhipov}},\ }\bibfield  {title} {\bibinfo {title} {Proceedings of the 43rd annual acm symposium on theory of computing, san jose}\ }(\bibinfo  {publisher} {ACM Press New York},\ \bibinfo {year} {2011})\BibitemShut {NoStop}%
\bibitem [{\citenamefont {Tamma}\ and\ \citenamefont {Laibacher}(2015)}]{tamma-multiboson}%
  \BibitemOpen
  \bibfield  {author} {\bibinfo {author} {\bibfnamefont {V.}~\bibnamefont {Tamma}}\ and\ \bibinfo {author} {\bibfnamefont {S.}~\bibnamefont {Laibacher}},\ }\bibfield  {title} {\bibinfo {title} {Multiboson correlation interferometry with arbitrary single-photon pure states},\ }\href {https://doi.org/10.1103/PhysRevLett.114.243601} {\bibfield  {journal} {\bibinfo  {journal} {Phys. Rev. Lett.}\ }\textbf {\bibinfo {volume} {114}},\ \bibinfo {pages} {243601} (\bibinfo {year} {2015})}\BibitemShut {NoStop}%
\bibitem [{\citenamefont {Laibacher}\ and\ \citenamefont {Tamma}(2015)}]{Tamma-Computational}%
  \BibitemOpen
  \bibfield  {author} {\bibinfo {author} {\bibfnamefont {S.}~\bibnamefont {Laibacher}}\ and\ \bibinfo {author} {\bibfnamefont {V.}~\bibnamefont {Tamma}},\ }\bibfield  {title} {\bibinfo {title} {From the physics to the computational complexity of multiboson correlation interference},\ }\href {https://doi.org/10.1103/PhysRevLett.115.243605} {\bibfield  {journal} {\bibinfo  {journal} {Phys. Rev. Lett.}\ }\textbf {\bibinfo {volume} {115}},\ \bibinfo {pages} {243605} (\bibinfo {year} {2015})}\BibitemShut {NoStop}%
\bibitem [{\citenamefont {Pont}\ \emph {et~al.}(2022)\citenamefont {Pont}, \citenamefont {Albiero}, \citenamefont {Thomas}, \citenamefont {Spagnolo}, \citenamefont {Ceccarelli}, \citenamefont {Corrielli}, \citenamefont {Brieussel}, \citenamefont {Somaschi}, \citenamefont {Huet}, \citenamefont {Harouri}, \citenamefont {Lema\^{\i}tre}, \citenamefont {Sagnes}, \citenamefont {Belabas}, \citenamefont {Sciarrino}, \citenamefont {Osellame}, \citenamefont {Senellart},\ and\ \citenamefont {Crespi}}]{SinglePhoton_Mathias}%
  \BibitemOpen
  \bibfield  {author} {\bibinfo {author} {\bibfnamefont {M.}~\bibnamefont {Pont}}, \bibinfo {author} {\bibfnamefont {R.}~\bibnamefont {Albiero}}, \bibinfo {author} {\bibfnamefont {S.~E.}\ \bibnamefont {Thomas}}, \bibinfo {author} {\bibfnamefont {N.}~\bibnamefont {Spagnolo}}, \bibinfo {author} {\bibfnamefont {F.}~\bibnamefont {Ceccarelli}}, \bibinfo {author} {\bibfnamefont {G.}~\bibnamefont {Corrielli}}, \bibinfo {author} {\bibfnamefont {A.}~\bibnamefont {Brieussel}}, \bibinfo {author} {\bibfnamefont {N.}~\bibnamefont {Somaschi}}, \bibinfo {author} {\bibfnamefont {H.}~\bibnamefont {Huet}}, \bibinfo {author} {\bibfnamefont {A.}~\bibnamefont {Harouri}}, \bibinfo {author} {\bibfnamefont {A.}~\bibnamefont {Lema\^{\i}tre}}, \bibinfo {author} {\bibfnamefont {I.}~\bibnamefont {Sagnes}}, \bibinfo {author} {\bibfnamefont {N.}~\bibnamefont {Belabas}}, \bibinfo {author} {\bibfnamefont {F.}~\bibnamefont {Sciarrino}}, \bibinfo {author} {\bibfnamefont {R.}~\bibnamefont {Osellame}}, \bibinfo {author} {\bibfnamefont
  {P.}~\bibnamefont {Senellart}},\ and\ \bibinfo {author} {\bibfnamefont {A.}~\bibnamefont {Crespi}},\ }\bibfield  {title} {\bibinfo {title} {Quantifying $n$-photon indistinguishability with a cyclic integrated interferometer},\ }\href {https://doi.org/10.1103/PhysRevX.12.031033} {\bibfield  {journal} {\bibinfo  {journal} {Phys. Rev. X}\ }\textbf {\bibinfo {volume} {12}},\ \bibinfo {pages} {031033} (\bibinfo {year} {2022})}\BibitemShut {NoStop}%
\bibitem [{\citenamefont {Ding}\ \emph {et~al.}(2025)\citenamefont {Ding}, \citenamefont {Guo}, \citenamefont {Xu}, \citenamefont {Liu}, \citenamefont {Zou}, \citenamefont {Zhao}, \citenamefont {Ge}, \citenamefont {Zhang}, \citenamefont {Liu}, \citenamefont {Wang}, \citenamefont {Chen}, \citenamefont {Wang}, \citenamefont {He}, \citenamefont {Huo}, \citenamefont {Lu},\ and\ \citenamefont {Pan}}]{Nature_SinglePhoton}%
  \BibitemOpen
  \bibfield  {author} {\bibinfo {author} {\bibfnamefont {X.}~\bibnamefont {Ding}}, \bibinfo {author} {\bibfnamefont {Y.-P.}\ \bibnamefont {Guo}}, \bibinfo {author} {\bibfnamefont {M.-C.}\ \bibnamefont {Xu}}, \bibinfo {author} {\bibfnamefont {R.-Z.}\ \bibnamefont {Liu}}, \bibinfo {author} {\bibfnamefont {G.-Y.}\ \bibnamefont {Zou}}, \bibinfo {author} {\bibfnamefont {J.-Y.}\ \bibnamefont {Zhao}}, \bibinfo {author} {\bibfnamefont {Z.-X.}\ \bibnamefont {Ge}}, \bibinfo {author} {\bibfnamefont {Q.-H.}\ \bibnamefont {Zhang}}, \bibinfo {author} {\bibfnamefont {H.-L.}\ \bibnamefont {Liu}}, \bibinfo {author} {\bibfnamefont {L.-J.}\ \bibnamefont {Wang}}, \bibinfo {author} {\bibfnamefont {M.-C.}\ \bibnamefont {Chen}}, \bibinfo {author} {\bibfnamefont {H.}~\bibnamefont {Wang}}, \bibinfo {author} {\bibfnamefont {Y.-M.}\ \bibnamefont {He}}, \bibinfo {author} {\bibfnamefont {Y.-H.}\ \bibnamefont {Huo}}, \bibinfo {author} {\bibfnamefont {C.-Y.}\ \bibnamefont {Lu}},\ and\ \bibinfo {author} {\bibfnamefont {J.-W.}\ \bibnamefont
  {Pan}},\ }\bibfield  {title} {\bibinfo {title} {High-efficiency single-photon source above the loss-tolerant threshold for efficient linear optical quantum computing},\ }\href {https://doi.org/10.1038/s41566-025-01639-8} {\bibfield  {journal} {\bibinfo  {journal} {Nat. Photonics}\ }\textbf {\bibinfo {volume} {19}},\ \bibinfo {pages} {387} (\bibinfo {year} {2025})}\BibitemShut {NoStop}%
\bibitem [{\citenamefont {Legero}\ \emph {et~al.}(2004)\citenamefont {Legero}, \citenamefont {Wilk}, \citenamefont {Hennrich}, \citenamefont {Rempe},\ and\ \citenamefont {Kuhn}}]{rempe-quanbeats}%
  \BibitemOpen
  \bibfield  {author} {\bibinfo {author} {\bibfnamefont {T.}~\bibnamefont {Legero}}, \bibinfo {author} {\bibfnamefont {T.}~\bibnamefont {Wilk}}, \bibinfo {author} {\bibfnamefont {M.}~\bibnamefont {Hennrich}}, \bibinfo {author} {\bibfnamefont {G.}~\bibnamefont {Rempe}},\ and\ \bibinfo {author} {\bibfnamefont {A.}~\bibnamefont {Kuhn}},\ }\bibfield  {title} {\bibinfo {title} {Quantum beat of two single photons},\ }\href {https://doi.org/10.1103/PhysRevLett.93.070503} {\bibfield  {journal} {\bibinfo  {journal} {Phys. Rev. Lett.}\ }\textbf {\bibinfo {volume} {93}},\ \bibinfo {pages} {070503} (\bibinfo {year} {2004})}\BibitemShut {NoStop}%
\bibitem [{\citenamefont {Wang}\ \emph {et~al.}(2018)\citenamefont {Wang}, \citenamefont {Jing}, \citenamefont {Sun}, \citenamefont {Yang}, \citenamefont {Yu}, \citenamefont {Tamma}, \citenamefont {Bao},\ and\ \citenamefont {Pan}}]{Tamma-three}%
  \BibitemOpen
  \bibfield  {author} {\bibinfo {author} {\bibfnamefont {X.-J.}\ \bibnamefont {Wang}}, \bibinfo {author} {\bibfnamefont {B.}~\bibnamefont {Jing}}, \bibinfo {author} {\bibfnamefont {P.-F.}\ \bibnamefont {Sun}}, \bibinfo {author} {\bibfnamefont {C.-W.}\ \bibnamefont {Yang}}, \bibinfo {author} {\bibfnamefont {Y.}~\bibnamefont {Yu}}, \bibinfo {author} {\bibfnamefont {V.}~\bibnamefont {Tamma}}, \bibinfo {author} {\bibfnamefont {X.-H.}\ \bibnamefont {Bao}},\ and\ \bibinfo {author} {\bibfnamefont {J.-W.}\ \bibnamefont {Pan}},\ }\bibfield  {title} {\bibinfo {title} {Experimental time-resolved interference with multiple photons of different colors},\ }\href {https://doi.org/10.1103/PhysRevLett.121.080501} {\bibfield  {journal} {\bibinfo  {journal} {Phys. Rev. Lett.}\ }\textbf {\bibinfo {volume} {121}},\ \bibinfo {pages} {080501} (\bibinfo {year} {2018})}\BibitemShut {NoStop}%
\bibitem [{\citenamefont {Orre}\ \emph {et~al.}(2019)\citenamefont {Orre}, \citenamefont {Goldschmidt}, \citenamefont {Deshpande}, \citenamefont {Gorshkov}, \citenamefont {Tamma}, \citenamefont {Hafezi},\ and\ \citenamefont {Mittal}}]{Tamma-freqresolved}%
  \BibitemOpen
  \bibfield  {author} {\bibinfo {author} {\bibfnamefont {V.~V.}\ \bibnamefont {Orre}}, \bibinfo {author} {\bibfnamefont {E.~A.}\ \bibnamefont {Goldschmidt}}, \bibinfo {author} {\bibfnamefont {A.}~\bibnamefont {Deshpande}}, \bibinfo {author} {\bibfnamefont {A.~V.}\ \bibnamefont {Gorshkov}}, \bibinfo {author} {\bibfnamefont {V.}~\bibnamefont {Tamma}}, \bibinfo {author} {\bibfnamefont {M.}~\bibnamefont {Hafezi}},\ and\ \bibinfo {author} {\bibfnamefont {S.}~\bibnamefont {Mittal}},\ }\bibfield  {title} {\bibinfo {title} {Interference of temporally distinguishable photons using frequency-resolved detection},\ }\href {https://doi.org/10.1103/PhysRevLett.123.123603} {\bibfield  {journal} {\bibinfo  {journal} {Phys. Rev. Lett.}\ }\textbf {\bibinfo {volume} {123}},\ \bibinfo {pages} {123603} (\bibinfo {year} {2019})}\BibitemShut {NoStop}%
\bibitem [{\citenamefont {Zou}\ \emph {et~al.}(1992)\citenamefont {Zou}, \citenamefont {Grayson},\ and\ \citenamefont {Mandel}}]{Zou_Grayson_Mandel_1992}%
  \BibitemOpen
  \bibfield  {author} {\bibinfo {author} {\bibfnamefont {X.~Y.}\ \bibnamefont {Zou}}, \bibinfo {author} {\bibfnamefont {T.~P.}\ \bibnamefont {Grayson}},\ and\ \bibinfo {author} {\bibfnamefont {L.}~\bibnamefont {Mandel}},\ }\bibfield  {title} {\bibinfo {title} {Observation of quantum interference effects in the frequency domain},\ }\href {https://doi.org/10.1103/PhysRevLett.69.3041} {\bibfield  {journal} {\bibinfo  {journal} {Phys. Rev. Lett.}\ }\textbf {\bibinfo {volume} {69}},\ \bibinfo {pages} {3041} (\bibinfo {year} {1992})}\BibitemShut {NoStop}%
\bibitem [{\citenamefont {Jin}\ \emph {et~al.}(2015)\citenamefont {Jin}, \citenamefont {Gerrits}, \citenamefont {Fujiwara}, \citenamefont {Wakabayashi}, \citenamefont {Yamashita}, \citenamefont {Miki}, \citenamefont {Terai}, \citenamefont {Shimizu}, \citenamefont {Takeoka},\ and\ \citenamefont {Sasaki}}]{NIST_Japan}%
  \BibitemOpen
  \bibfield  {author} {\bibinfo {author} {\bibfnamefont {R.-B.}\ \bibnamefont {Jin}}, \bibinfo {author} {\bibfnamefont {T.}~\bibnamefont {Gerrits}}, \bibinfo {author} {\bibfnamefont {M.}~\bibnamefont {Fujiwara}}, \bibinfo {author} {\bibfnamefont {R.}~\bibnamefont {Wakabayashi}}, \bibinfo {author} {\bibfnamefont {T.}~\bibnamefont {Yamashita}}, \bibinfo {author} {\bibfnamefont {S.}~\bibnamefont {Miki}}, \bibinfo {author} {\bibfnamefont {H.}~\bibnamefont {Terai}}, \bibinfo {author} {\bibfnamefont {R.}~\bibnamefont {Shimizu}}, \bibinfo {author} {\bibfnamefont {M.}~\bibnamefont {Takeoka}},\ and\ \bibinfo {author} {\bibfnamefont {M.}~\bibnamefont {Sasaki}},\ }\bibfield  {title} {\bibinfo {title} {Spectrally resolved hong-ou-mandel interference between independent photon sources},\ }\href {https://doi.org/10.1364/OE.23.028836} {\bibfield  {journal} {\bibinfo  {journal} {Opt. Express}\ }\textbf {\bibinfo {volume} {23}},\ \bibinfo {pages} {28836} (\bibinfo {year} {2015})}\BibitemShut {NoStop}%
\bibitem [{\citenamefont {Prakash}\ \emph {et~al.}(2021)\citenamefont {Prakash}, \citenamefont {Sierant},\ and\ \citenamefont {Mitchell}}]{Mitchell}%
  \BibitemOpen
  \bibfield  {author} {\bibinfo {author} {\bibfnamefont {V.}~\bibnamefont {Prakash}}, \bibinfo {author} {\bibfnamefont {A.}~\bibnamefont {Sierant}},\ and\ \bibinfo {author} {\bibfnamefont {M.~W.}\ \bibnamefont {Mitchell}},\ }\bibfield  {title} {\bibinfo {title} {Autoheterodyne characterization of narrow-band photon pairs},\ }\href {https://doi.org/10.1103/PhysRevLett.127.043601} {\bibfield  {journal} {\bibinfo  {journal} {Phys. Rev. Lett.}\ }\textbf {\bibinfo {volume} {127}},\ \bibinfo {pages} {043601} (\bibinfo {year} {2021})}\BibitemShut {NoStop}%
\bibitem [{\citenamefont {Drexler}\ \emph {et~al.}(2008)\citenamefont {Drexler}, \citenamefont {Fujimoto}, \citenamefont {Drexler},\ and\ \citenamefont {Fujimoto}}]{OCT}%
  \BibitemOpen
  \bibfield  {author} {\bibinfo {author} {\bibfnamefont {W.}~\bibnamefont {Drexler}}, \bibinfo {author} {\bibfnamefont {J.~G.}\ \bibnamefont {Fujimoto}}, \bibinfo {author} {\bibfnamefont {W.}~\bibnamefont {Drexler}},\ and\ \bibinfo {author} {\bibfnamefont {J.~G.}\ \bibnamefont {Fujimoto}},\ }\href@noop {} {\emph {\bibinfo {title} {Optical Coherence Tomography: Technology and Applications}}},\ \bibinfo {edition} {1st}\ ed.\ (\bibinfo  {publisher} {Springer-Verlag},\ \bibinfo {address} {Berlin, Heidelberg},\ \bibinfo {year} {2008})\BibitemShut {NoStop}%
\bibitem [{\citenamefont {Yao}\ and\ \citenamefont {Padgett}(2011)}]{OAM-Padgett-2011}%
  \BibitemOpen
  \bibfield  {author} {\bibinfo {author} {\bibfnamefont {A.~M.}\ \bibnamefont {Yao}}\ and\ \bibinfo {author} {\bibfnamefont {M.~J.}\ \bibnamefont {Padgett}},\ }\bibfield  {title} {\bibinfo {title} {Orbital angular momentum: origins, behavior and applications},\ }\href {https://doi.org/10.1364/AOP.3.000161} {\bibfield  {journal} {\bibinfo  {journal} {Adv. Opt. Photon.}\ }\textbf {\bibinfo {volume} {3}},\ \bibinfo {pages} {161} (\bibinfo {year} {2011})}\BibitemShut {NoStop}%
\bibitem [{\citenamefont {Padgett}(2017)}]{OAM-Padgett-2017}%
  \BibitemOpen
  \bibfield  {author} {\bibinfo {author} {\bibfnamefont {M.~J.}\ \bibnamefont {Padgett}},\ }\bibfield  {title} {\bibinfo {title} {Orbital angular momentum 25 years on [invited]},\ }\href {https://doi.org/10.1364/OE.25.011265} {\bibfield  {journal} {\bibinfo  {journal} {Opt. Express}\ }\textbf {\bibinfo {volume} {25}},\ \bibinfo {pages} {11265} (\bibinfo {year} {2017})}\BibitemShut {NoStop}%
\bibitem [{\citenamefont {Kim}\ \emph {et~al.}(2006)\citenamefont {Kim}, \citenamefont {Kwon}, \citenamefont {Kim},\ and\ \citenamefont {Kim}}]{Kim-spatial-HOM-fermion-boson}%
  \BibitemOpen
  \bibfield  {author} {\bibinfo {author} {\bibfnamefont {H.}~\bibnamefont {Kim}}, \bibinfo {author} {\bibfnamefont {O.}~\bibnamefont {Kwon}}, \bibinfo {author} {\bibfnamefont {W.}~\bibnamefont {Kim}},\ and\ \bibinfo {author} {\bibfnamefont {T.}~\bibnamefont {Kim}},\ }\bibfield  {title} {\bibinfo {title} {Spatial two-photon interference in a hong-ou-mandel interferometer},\ }\href {https://doi.org/10.1103/PhysRevA.73.023820} {\bibfield  {journal} {\bibinfo  {journal} {Phys. Rev. A}\ }\textbf {\bibinfo {volume} {73}},\ \bibinfo {pages} {023820} (\bibinfo {year} {2006})}\BibitemShut {NoStop}%
\bibitem [{\citenamefont {Lee}\ and\ \citenamefont {van Exter}(2006)}]{Lee-spatial-label}%
  \BibitemOpen
  \bibfield  {author} {\bibinfo {author} {\bibfnamefont {P.~S.~K.}\ \bibnamefont {Lee}}\ and\ \bibinfo {author} {\bibfnamefont {M.~P.}\ \bibnamefont {van Exter}},\ }\bibfield  {title} {\bibinfo {title} {Spatial labeling in a two-photon interferometer},\ }\href {https://doi.org/10.1103/PhysRevA.73.063827} {\bibfield  {journal} {\bibinfo  {journal} {Phys. Rev. A}\ }\textbf {\bibinfo {volume} {73}},\ \bibinfo {pages} {063827} (\bibinfo {year} {2006})}\BibitemShut {NoStop}%
\bibitem [{\citenamefont {Guo}\ \emph {et~al.}(2025)\citenamefont {Guo}, \citenamefont {Chen}, \citenamefont {Maggio}, \citenamefont {Wu}, \citenamefont {Liu}, \citenamefont {Tamma},\ and\ \citenamefont {Fan}}]{QuantumBeat_Momentum}%
  \BibitemOpen
  \bibfield  {author} {\bibinfo {author} {\bibfnamefont {B.}~\bibnamefont {Guo}}, \bibinfo {author} {\bibfnamefont {Z.}~\bibnamefont {Chen}}, \bibinfo {author} {\bibfnamefont {L.}~\bibnamefont {Maggio}}, \bibinfo {author} {\bibfnamefont {W.}~\bibnamefont {Wu}}, \bibinfo {author} {\bibfnamefont {S.}~\bibnamefont {Liu}}, \bibinfo {author} {\bibfnamefont {V.}~\bibnamefont {Tamma}},\ and\ \bibinfo {author} {\bibfnamefont {J.}~\bibnamefont {Fan}},\ }\bibfield  {title} {\bibinfo {title} {Quantum beat of two single photons in the transverse momentum space},\ }\href {https://doi.org/10.1103/wv2d-5z4g} {\bibfield  {journal} {\bibinfo  {journal} {Phys. Rev. A}\ }\textbf {\bibinfo {volume} {112}},\ \bibinfo {pages} {013719} (\bibinfo {year} {2025})}\BibitemShut {NoStop}%
\bibitem [{\citenamefont {Ghosh}\ and\ \citenamefont {Mandel}(1987)}]{Ghosh_Mandel}%
  \BibitemOpen
  \bibfield  {author} {\bibinfo {author} {\bibfnamefont {R.}~\bibnamefont {Ghosh}}\ and\ \bibinfo {author} {\bibfnamefont {L.}~\bibnamefont {Mandel}},\ }\bibfield  {title} {\bibinfo {title} {Observation of nonclassical effects in the interference of two photons},\ }\href {https://doi.org/10.1103/PhysRevLett.59.1903} {\bibfield  {journal} {\bibinfo  {journal} {Phys. Rev. Lett.}\ }\textbf {\bibinfo {volume} {59}},\ \bibinfo {pages} {1903} (\bibinfo {year} {1987})}\BibitemShut {NoStop}%
\bibitem [{\citenamefont {Ou}\ and\ \citenamefont {Mandel}(1989)}]{Ou_Mandel_1989}%
  \BibitemOpen
  \bibfield  {author} {\bibinfo {author} {\bibfnamefont {Z.~Y.}\ \bibnamefont {Ou}}\ and\ \bibinfo {author} {\bibfnamefont {L.}~\bibnamefont {Mandel}},\ }\bibfield  {title} {\bibinfo {title} {Further evidence of nonclassical behavior in optical interference},\ }\href {https://doi.org/10.1103/PhysRevLett.62.2941} {\bibfield  {journal} {\bibinfo  {journal} {Phys. Rev. Lett.}\ }\textbf {\bibinfo {volume} {62}},\ \bibinfo {pages} {2941} (\bibinfo {year} {1989})}\BibitemShut {NoStop}%
\bibitem [{\citenamefont {Glauber}(1963)}]{glauber-1963}%
  \BibitemOpen
  \bibfield  {author} {\bibinfo {author} {\bibfnamefont {R.~J.}\ \bibnamefont {Glauber}},\ }\bibfield  {title} {\bibinfo {title} {The quantum theory of optical coherence},\ }\href {https://doi.org/10.1103/PhysRev.130.2529} {\bibfield  {journal} {\bibinfo  {journal} {Phys. Rev.}\ }\textbf {\bibinfo {volume} {130}},\ \bibinfo {pages} {2529} (\bibinfo {year} {1963})}\BibitemShut {NoStop}%
\bibitem [{\citenamefont {Valencia}\ \emph {et~al.}(2002)\citenamefont {Valencia}, \citenamefont {Chekhova}, \citenamefont {Trifonov},\ and\ \citenamefont {Shih}}]{Masha-biphoton}%
  \BibitemOpen
  \bibfield  {author} {\bibinfo {author} {\bibfnamefont {A.}~\bibnamefont {Valencia}}, \bibinfo {author} {\bibfnamefont {M.~V.}\ \bibnamefont {Chekhova}}, \bibinfo {author} {\bibfnamefont {A.}~\bibnamefont {Trifonov}},\ and\ \bibinfo {author} {\bibfnamefont {Y.}~\bibnamefont {Shih}},\ }\bibfield  {title} {\bibinfo {title} {Entangled two-photon wave packet in a dispersive medium},\ }\href {https://doi.org/10.1103/PhysRevLett.88.183601} {\bibfield  {journal} {\bibinfo  {journal} {Phys. Rev. Lett.}\ }\textbf {\bibinfo {volume} {88}},\ \bibinfo {pages} {183601} (\bibinfo {year} {2002})}\BibitemShut {NoStop}%
\bibitem [{\citenamefont {Mandel}(1962)}]{Mandel-alford-gold:62}%
  \BibitemOpen
  \bibfield  {author} {\bibinfo {author} {\bibfnamefont {L.}~\bibnamefont {Mandel}},\ }\bibfield  {title} {\bibinfo {title} {Interference and the alford and gold effect},\ }\href {https://doi.org/10.1364/JOSA.52.001335} {\bibfield  {journal} {\bibinfo  {journal} {J. Opt. Soc. Am.}\ }\textbf {\bibinfo {volume} {52}},\ \bibinfo {pages} {1335} (\bibinfo {year} {1962})}\BibitemShut {NoStop}%
\bibitem [{\citenamefont {Salazar-Serrano}\ \emph {et~al.}(2014)\citenamefont {Salazar-Serrano}, \citenamefont {Valencia},\ and\ \citenamefont {Torres}}]{Salazar-Serrano:14}%
  \BibitemOpen
  \bibfield  {author} {\bibinfo {author} {\bibfnamefont {L.~J.}\ \bibnamefont {Salazar-Serrano}}, \bibinfo {author} {\bibfnamefont {A.}~\bibnamefont {Valencia}},\ and\ \bibinfo {author} {\bibfnamefont {J.~P.}\ \bibnamefont {Torres}},\ }\bibfield  {title} {\bibinfo {title} {Observation of spectral interference for any path difference in an interferometer},\ }\href@noop {} {\bibfield  {journal} {\bibinfo  {journal} {Opt. Lett.}\ }\textbf {\bibinfo {volume} {39}},\ \bibinfo {pages} {4478} (\bibinfo {year} {2014})}\BibitemShut {NoStop}%
\bibitem [{\citenamefont {Flórez}\ \emph {et~al.}(2016)\citenamefont {Flórez}, \citenamefont {Álvarez}, \citenamefont {Calderón-Losada}, \citenamefont {Salazar-Serrano},\ and\ \citenamefont {Valencia}}]{Flórez_2016}%
  \BibitemOpen
  \bibfield  {author} {\bibinfo {author} {\bibfnamefont {J.}~\bibnamefont {Flórez}}, \bibinfo {author} {\bibfnamefont {J.-R.}\ \bibnamefont {Álvarez}}, \bibinfo {author} {\bibfnamefont {O.}~\bibnamefont {Calderón-Losada}}, \bibinfo {author} {\bibfnamefont {L.~J.}\ \bibnamefont {Salazar-Serrano}},\ and\ \bibinfo {author} {\bibfnamefont {A.}~\bibnamefont {Valencia}},\ }\bibfield  {title} {\bibinfo {title} {Interference of two pulse-like spatial beams with arbitrary transverse separation},\ }\href {https://doi.org/10.1088/2040-8978/18/12/125201} {\bibfield  {journal} {\bibinfo  {journal} {J. Opt.}\ }\textbf {\bibinfo {volume} {18}},\ \bibinfo {pages} {125201} (\bibinfo {year} {2016})}\BibitemShut {NoStop}%
\bibitem [{\citenamefont {T.~Legero}(2003)}]{leggero-teoria}%
  \BibitemOpen
  \bibfield  {author} {\bibinfo {author} {\bibfnamefont {A.~K. . G.~R.}\ \bibnamefont {T.~Legero}, \bibfnamefont {T.~Wilk}},\ }\bibfield  {title} {\bibinfo {title} {Time-resolved two-photon quantum interference},\ }\href {https://doi.org/10.1007/s00340-003-1337-x} {\bibfield  {journal} {\bibinfo  {journal} {Appl. Phys. B}\ }\textbf {\bibinfo {volume} {77}},\ \bibinfo {pages} {797} (\bibinfo {year} {2003})}\BibitemShut {NoStop}%
\bibitem [{dra()}]{drawingpage}%
  \BibitemOpen
  \href@noop {} {\bibinfo {title} {Optical components pack v1}},\ \bibinfo {howpublished} {\url{https://ryomizutagraphics.gumroad.com/l/OpticalComponentsV1}}\BibitemShut {NoStop}%
\bibitem [{\citenamefont {Salazar-Serrano}\ \emph {et~al.}(2015)\citenamefont {Salazar-Serrano}, \citenamefont {Valencia},\ and\ \citenamefont {Torres}}]{Salazar-PTBD}%
  \BibitemOpen
  \bibfield  {author} {\bibinfo {author} {\bibfnamefont {L.~J.}\ \bibnamefont {Salazar-Serrano}}, \bibinfo {author} {\bibfnamefont {A.}~\bibnamefont {Valencia}},\ and\ \bibinfo {author} {\bibfnamefont {J.~P.}\ \bibnamefont {Torres}},\ }\bibfield  {title} {\bibinfo {title} {Tunable beam displacer},\ }\href {https://doi.org/10.1063/1.4914834} {\bibfield  {journal} {\bibinfo  {journal} {Rev. Sci. Instrum.}\ }\textbf {\bibinfo {volume} {86}},\ \bibinfo {pages} {033109} (\bibinfo {year} {2015})}\BibitemShut {NoStop}%
\bibitem [{\citenamefont {Perez-Leija}\ \emph {et~al.}(2018)\citenamefont {Perez-Leija}, \citenamefont {Guzm{\'a}n-Silva}, \citenamefont {Le{\'o}n-Montiel}, \citenamefont {Gr{\"a}fe}, \citenamefont {Heinrich}, \citenamefont {Moya-Cessa}, \citenamefont {Busch},\ and\ \citenamefont {Szameit}}]{Fidelity_QuantumCoherence}%
  \BibitemOpen
  \bibfield  {author} {\bibinfo {author} {\bibfnamefont {A.}~\bibnamefont {Perez-Leija}}, \bibinfo {author} {\bibfnamefont {D.}~\bibnamefont {Guzm{\'a}n-Silva}}, \bibinfo {author} {\bibfnamefont {R.~d.~J.}\ \bibnamefont {Le{\'o}n-Montiel}}, \bibinfo {author} {\bibfnamefont {M.}~\bibnamefont {Gr{\"a}fe}}, \bibinfo {author} {\bibfnamefont {M.}~\bibnamefont {Heinrich}}, \bibinfo {author} {\bibfnamefont {H.}~\bibnamefont {Moya-Cessa}}, \bibinfo {author} {\bibfnamefont {K.}~\bibnamefont {Busch}},\ and\ \bibinfo {author} {\bibfnamefont {A.}~\bibnamefont {Szameit}},\ }\bibfield  {title} {\bibinfo {title} {Endurance of quantum coherence due to particle indistinguishability in noisy quantum networks},\ }\href@noop {} {\bibfield  {journal} {\bibinfo  {journal} {npj Quantum Inf.}\ }\textbf {\bibinfo {volume} {4}},\ \bibinfo {pages} {45} (\bibinfo {year} {2018})}\BibitemShut {NoStop}%
\bibitem [{\citenamefont {Peruzzo}\ \emph {et~al.}(2010)\citenamefont {Peruzzo}, \citenamefont {Lobino}, \citenamefont {Matthews}, \citenamefont {Matsuda}, \citenamefont {Politi}, \citenamefont {Poulios}, \citenamefont {Zhou}, \citenamefont {Lahini}, \citenamefont {Ismail}, \citenamefont {Wörhoff}, \citenamefont {Bromberg}, \citenamefont {Silberberg}, \citenamefont {Thompson},\ and\ \citenamefont {OBrien}}]{Fidelity_QuantumWalks}%
  \BibitemOpen
  \bibfield  {author} {\bibinfo {author} {\bibfnamefont {A.}~\bibnamefont {Peruzzo}}, \bibinfo {author} {\bibfnamefont {M.}~\bibnamefont {Lobino}}, \bibinfo {author} {\bibfnamefont {J.~C.~F.}\ \bibnamefont {Matthews}}, \bibinfo {author} {\bibfnamefont {N.}~\bibnamefont {Matsuda}}, \bibinfo {author} {\bibfnamefont {A.}~\bibnamefont {Politi}}, \bibinfo {author} {\bibfnamefont {K.}~\bibnamefont {Poulios}}, \bibinfo {author} {\bibfnamefont {X.-Q.}\ \bibnamefont {Zhou}}, \bibinfo {author} {\bibfnamefont {Y.}~\bibnamefont {Lahini}}, \bibinfo {author} {\bibfnamefont {N.}~\bibnamefont {Ismail}}, \bibinfo {author} {\bibfnamefont {K.}~\bibnamefont {Wörhoff}}, \bibinfo {author} {\bibfnamefont {Y.}~\bibnamefont {Bromberg}}, \bibinfo {author} {\bibfnamefont {Y.}~\bibnamefont {Silberberg}}, \bibinfo {author} {\bibfnamefont {M.~G.}\ \bibnamefont {Thompson}},\ and\ \bibinfo {author} {\bibfnamefont {J.~L.}\ \bibnamefont {OBrien}},\ }\bibfield  {title} {\bibinfo {title} {Quantum walks of correlated photons},\ }\href@noop {}
  {\bibfield  {journal} {\bibinfo  {journal} {Science}\ }\textbf {\bibinfo {volume} {329}},\ \bibinfo {pages} {1500} (\bibinfo {year} {2010})}\BibitemShut {NoStop}%
\end{thebibliography}%

\end{document}